\documentclass[reprint,twocolumn,showpacs,showkeys,superscriptaddress,nofootinbib]{revtex4-1}
\usepackage{graphicx,amssymb,amsmath,multirow}

\begin{document}
%-----------------------------------------------------------------
\title{Anisotropic flow of the fireball fed by hard partons}
\author{Martin Schulc}\email{martin.schulc@fjfi.cvut.cz}
\affiliation{Czech Technical University in Prague, FNSPE, CZ 11519 Prague 1, Czech Republic}
\author{Boris~Tom\'a\v{s}ik}\email{boris.tomasik@umb.sk}
\affiliation{Univerzita Mateja Bela, SK 97401 Bansk\'a Bystrica, Slovakia \\ and 
Czech Technical University in Prague, FNSPE, CZ 11519 Prague 1, Czech Republic
}

%-----------------------------------------------------------------------
\begin{abstract}
In nuclear collisions at highest accessible LHC energies, often more than one dijet pairs
deposit momentum into the deconfined expanding medium. With the help of 3+1 dimensional
relativistic hydrodynamic simulation we show that this leads to measurable contribution 
to the anisotropy of collective transverse expansion. Hard partons generate streams in plasma
which merge if they come close to each other. This mechanism correlates the resulting
contribution to flow anisotropy with the fireball geometry and causes an increase of the 
elliptic flow in non-central collisions. 
\end{abstract}
\date{December 3, 2014}
%{\sf File name:~\jobname}
\pacs{25.75.-q, 25.75.Ld
%21.65.-f % Nuclear matter
}
\keywords{heavy-ion collision, anisotropic flow, hydrodynamic simulation, jets}
\maketitle

Study of the properties of the hottest matter ever created in laboratory is in the focus 
of the heavy-ion programme at the LHC. From data on jet quenching we know that the 
created matter is in deconfined state. Currently, the focus is on studying the properties of such 
deconfined strongly interacting matter. Comparisons of hydrodynamic simulations with the 
measured data aim at extracting the transport coefficients, mainly the viscosity. 

Due to transverse expansion of the created hot matter, hadronic transverse momentum spectra
show a blue-shift. The blue-shift varies azimuthally. This indicates the modulation of
the transverse expansion velocity as a function of the azimuthal angle. Such a modulation 
appears naturally in non-central collisions due to azimuthally asymmetric shape of the 
initial overlap region. However, a more detailed analysis reveals azimuthal 
anisotropies in every event which are causally linked to to fluctuations in the initial state
\cite{Alver-Roland,Schenke-fluctuations,Holopainen,Wiedemann-Floerchinger,Qian:2013nba,
Heinz:2013wva}. 
As these fluctuations are propagated within the (weakly) viscous relativistic fluid, dedicated 
simulation could put relevant limits on the transport properties of the deconfined matter
\cite{Schenke-fluctuations}.
This is the standard approach which is being used in present investigations: by selecting a
set of initial conditions and tuning the values of viscosities one tries to find such a setting 
of hydrodynamic simulations which reproduces as many features of data as possible. The 
data today are very rich with a few orders of azimuthal anisotropies for identified species, 
many kinds of correlations, everything measured in various centrality classes 
\cite{ALICEvn,Abelev:2014pua,Aad:2014vba,Chatrchyan:2013kba,CMSvn}.

In this paper we point out another source of spectral azimuthal anisotropy. It cannot 
be put into the family of models where initial conditions are exclusively responsible for the
anisotropy. At the LHC, jets are no longer such a rare probe. They are produced in initial 
hard scattering together with copious minijets and propagate through the deconfined 
medium. It is known that quark-gluon plasma quenches a large part---if not all---of 
the energy and momentum of the hard partons which might become jets. The momentum 
deposition from the partons into medium induces collective effects 
\cite{Satarov:2005mv, CasalderreySolana:2004qm, Koch:2005sx, Ruppert:2005uz, Renk:2005si,
Neufeld:2008hs, Neufeld:2010tz, Betz:2008ka, Betz:2010qh,Renk:2013pua}
and owing to momentum conservation there must be net flow. Recently in 
\cite{Tachibana} the 
response of medium to one very energetic dijet was simulated in 3+1D hydrodynamics. 
In \cite{Noronha} the generation of elliptic and triangular 
flow due to hard partons within a 2+1D model was simulated. 
The introduction of jets, however, breaks 
longitudinal boost invariance which is implicitly assumed in a 2+1D simulation. The influence 
of jets on the evolution in central collisions was investigated in a 1+1D approach also 
in \cite{Florchinger,Zapp}. Here we present results from our three-dimensional ideal 
hydrodynamic simulation with realistic multiplicity distribution of hard partons. 

In \cite{Tomasik:2011xn} it was shown with a help of a toy model that if there are a few pairs of 
minijets within one event, the wakes which they deposit may influence each other 
and so lead to elliptic flow anisotropy correlated with the reaction plane. 
Later in \cite{schulc} we have shown 
that the concept of two merging wakes that follow as one stream is reproduced in ideal 
hydrodynamics in a static medium. Here we apply these ideas in three-dimensional 
simulations of an expanding fireball motivated by realistic collision dynamics. 

We present results on first to fourth order flow anisotropies 
in central and non-central collisions. Hard partons depositing 
momentum themselves are capable of generating 
$v_2$ of the order 0.015 in ultra-central collisions at the LHC. It is important that in 
non-central collisions their  contribution is correlated with fireball geometry. We show that 
they contribute considerably to the observed anisotropy of hadronic spectra. 

Higher harmonics of azimuthal anisotropy from hydrodynamically expanding fireball
complemented with jets were recently calculated also within the HYDJET++ model 
\cite{Bravina}. However, that model consists  of two \emph{independent} parts: the 
soft production is modelled by parametrisation of hadron emission while the hard part is 
simulated by separate Monte Carlo model. In contrast to that, we focus on the 
\emph{interplay} of the fluid and the jets. We study how the fluid behaves when it is 
stimulated by jets in addition to expansion due to pressure gradients. 

We perform event-by-event hydrodynamic simulations. Our model is three-dimensional, based 
on ideal hydrodynamics and uses the SHASTA algorithm \cite{Boris,DeVore} to deal with shock 
fronts. For each event the initial conditions are first constructed smooth according to
the optical Glauber prescription. Transverse profile of the energy density at 
impact parameter $b$ is characterised by 
\begin{equation}
W(x,y;b) = (1-\alpha) n_w(x,y;b) + \alpha n_{\mathrm{bin}}(x,y;b)
\end{equation}
where $n_w$ and $n_{\mathrm{bin}}$ are the numbers of wounded nucleons and binary collisions
at given transverse position $(x,y)$ and the coefficient $\alpha$ is set to 0.16.
The nucleon-nucleon cross-section for Glauber calculation at 
$\sqrt{s_{NN}} = 5.5$~TeV is set to 62~mb. By choosing a smooth transverse profile 
with no event-by-event fluctuations we can later be sure that any anisotropic flow in 
addition to the event-averaged one is due to the contribution of hard partons. We can
thus better estimate their contribution. For the 3+1D hydrodynamic simulation, initial 
profile in space-time rapidity $\eta_s = \frac{1}{2}\log((t+z)/(t-z))$ is given by 
\begin{equation}
H(\eta_s) = \exp\left (  
- \frac{\left ( |\eta_s| - \eta_{\mathrm{flat}}/2\right )^2}{2\sigma_\eta^2} \right )\,
\theta\left (  |\eta_s| - \eta_{\mathrm{flat}}/2 \right )\,  .
\end{equation}
We chose $\eta_{\mathrm{flat}}=10$  \cite{Schenke2} 
and $\sigma_\eta = 0.5$.
The initial energy density
then follows the distribution
\begin{equation}
\epsilon(x,y,\eta_s;b) = \epsilon_0 \, \frac{W(x,y;b)}{W(0,0;0)}\, H(\eta_s)\,  .
\end{equation}
We choose $\epsilon_0 = 60$~GeV/fm$^3$ for the initial longitudinal 
proper time $\tau = 0.55$~fm/$c$. 

For the hydrodynamic evolution we have taken lattice-inspired 
Equation of State from \cite{Petreczky}. 

Momentum feeding from hard partons into medium is implemented via source 
terms in the energy-momentum conservation equation
\begin{equation}
\partial_\mu T^{\mu\nu} = J^\nu,
\end{equation}
where the source term $J^\nu$ stands for the rate of energy-momemtum loss of hard parton
\cite{Betz:2008ka,Betz:2010qh}
\begin{equation}
J^\nu = -\sum_i \int_{\tau_{i,i}}^{\tau_{f,i}}d\tau\, \frac{dP_i^\nu}{d\tau}\, 
\delta^{(4)}\left (  x^\mu - x_{\mathrm{jet},i}^\mu  \right ),
\end{equation}
where $P_i^\mu$ and $x_{\mathrm{jet},i}^\nu$ denote momentum and position of 
the $i$-th hard parton, respectively. The sign in front of the summation reflects the fact that 
the change of momentum of the medium is opposite to the momentum change of the hard 
parton. Integration runs over the whole lifetime 
of $i$-th parton until its energy is fully  quenched and the summation goes over 
all hard partons of the event. The microscopic picture of how momentum is transferred
from the parton into medium is being investigated \cite{Neufeld:2008,Neufeld-sourceterm} 
but not yet fully understood 
at an applicable level. We thus introduce spatial region over which the momentum is 
initially distributed in a non-covariant implementation of the source term
\begin{equation}
J^\nu = -\sum_i \frac{1}{(2\, \pi\, \sigma_i^2)^{\frac{3}{2}}} \, \exp \left (
- \frac{\left ( \vec x - \vec x_{\mathrm{jet},i} \right )^2 }{2\, \sigma_i^2} \right )\, 
\left ( \frac{dE_i}{dt}, \frac{d\vec P_i}{dt} \right )
\end{equation}
with $\sigma = 0.3$~fm. Partons are assumed to have mass 0.3~GeV when momentum 
loss is determined from the energy loss. 

Parton energy loss depends on the density of the medium. The exact form of
this dependence is not known, yet \cite{Betz_Gyulassy, renk_Eloss_scaling}. 
Here we assume that it scales with entropy density $s$ \cite{Betz:2010tb}. 
The scaling relation is thus
\begin{equation}
\frac{dE}{dx} = \left . \frac{dE}{dx}\right |_0 \,  \frac{s}{s_0} 
\end{equation}
with $s_0$ corresponding to energy density 20.0~GeV/fm$^3$ ($T=324$~MeV and $s=78.2$/fm$^{3}$)
For $dE/dx|_0$ we usually choose values 4 and 7~GeV/fm. 

For the production of hard partons we take the parametrisation of gluon cross-section per
nucleon-nucleon pair in nucleus-nucleus collisions
\begin{equation}
E \frac{d\sigma_{NN}}{d^3p} = \frac{1}{2\pi}\, \frac{1}{p_t}\, \frac{d\sigma_{NN}}{dp_t\, dy}
= \frac{B}{(1+p_t/p_0)^n}
\end{equation}
where for the energy $\sqrt{s_{NN}} =  5.5$~TeV we have $B = 14.7$~mb/GeV$^2$, 
$p_0 = 6$~GeV and $n = 9.5$. The distribution of hard parton pairs in transverse plane 
scales with the number of binary collisions. The pairs have 
balanced transverse momentum. For the presented results we generated dijet pairs with $p_t$ 
above 3~GeV. 

We chose to make simulations for the collision energy of $\sqrt{s_{NN}} =  5.5$~TeV
for which one of us published the educated guess that the effects of momentum 
deposition on flow anisotropy should be measurable \cite{Tomasik:2011xn}. 
This can also be regarded as a prediction for future LHC run. 

Freeze-out is handled by the Cooper-Frye prescription \cite{CooperFrye} 
on the hypersurface given 
by $T = 150$~MeV. We use the THERMINATOR2 package
\cite{THERM2} to generate hadrons on the obtained hypersurface and evaluate results. 

For Pb+Pb collision at full LHC energy $\sqrt{s_{NN}} = 5.5$~TeV we simulate sets of events 
in three centrality classes. In order to establish the effect on anisotropic flow due to 
hard partons we analyze two central classes of events: one corresponding to 0--2.5\%
of centrality distribution and one where we strictly set the impact parameter $b=0$~fm. 
In order to see the contribution of our mechanism in non-central collisions, we also simulate 
a set of 30--40\% centrality class. 

For each setting we generate 100 hydrodynamic events. 
On top of that we run on each obtained hypersurface five times the THERMINATOR2 freeze-out 
procedure and thus we quintuple the number of events in the analysis. 
Resonance decays are included. 
We obtain the anisotropic flow parameters $v_1$, $v_2$, $v_3$,
$v_4$ for charged hadrons by the two-particle cumulant method. 
Recall that we analyse hadrons coming from the bulk freeze-out of hot matter with 
collective flow influenced by hard partons. All anisotropies in hadronic distributions 
are due to anisotropic collective expansion. 

\begin{figure}[ht!]
\includegraphics[height=0.87\textheight]{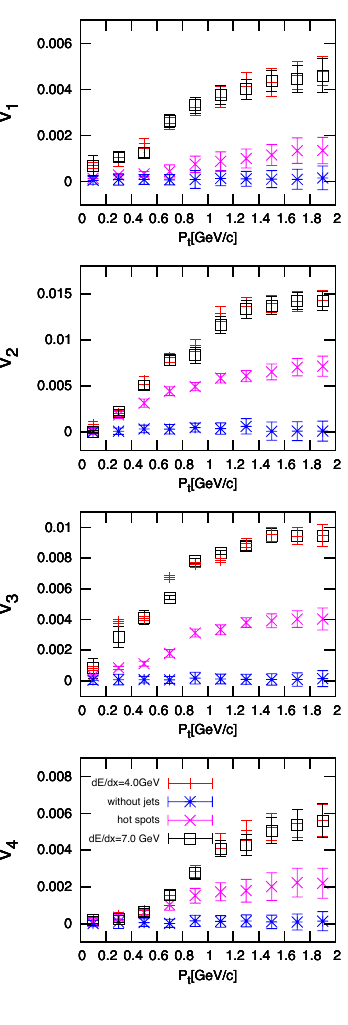}
\caption{\label{f:b0}%
Parameters $v_n$ from collisions  at $b=0$ for charged 
hadrons. Different symbols represent: energy loss of hard 
parton $dE/dx|_0 = 4$~GeV/fm (red --), 7~GeV/fm
(black $\square$), scenario with only hot spots in initial 
conditions (purple $\times$), scenario with smooth initial conditions (blue *). 
}
\end{figure}

We first investigate the size of generated anisotropy of momentum distribution in 
ultra-central collisions ($b=0$). Results are shown in Figure~\ref{f:b0}.
%%%%%%%%%%%%%%%%%%%%%%%%%%%%%%%%%%%%%%%%%%%%%%%%%%
%%%%%%%%%%%%%%%%%%%%%%%%%%%%%%%%%%%%%%%%%%%%%%%%%%%
Two values for the energy loss are tested: 
$dE/dx|_0 = 4$~GeV/fm and 7~GeV/fm. As a benchmark test we also evaluate the 
$v_n$'s from simulation with no hard partons and no fluctuations and show that they are 
consistent with 0. The results are also compared with simulations where hot spots
were superimposed on the smooth energy density profile. There are as many 
hot spots as there would be hard partons. These are regions where we deposit 
the same amount of energy that a hard parton would carry if it was produced
there. In contrast to hard partons, in hot spots the energy is included in the initial 
conditions and not released over finite time interval. Also, in a hot-spots scenario  
no momentum is deposited. The comparison in Fig.~\ref{f:b0} shows that momentum 
deposition is important. Fluctuations in the initial conditions by themselves 
are not able to generate the same flow anisotropies as wakes with streams 
induced by hard partons. 

It is somewhat puzzling why there is no difference in results between the two scenarios 
which differ in the value of the energy loss. Choosing higher value of 
$dE/dx$ causes that the 
partons loose their momentum faster, but it is the same total amount of momentum 
that is deposited into the fluid. In fact, most of them have rather low $p_t$ 
and thus are quenched early. 

\begin{figure}[t]
\includegraphics[width=0.44\textwidth]{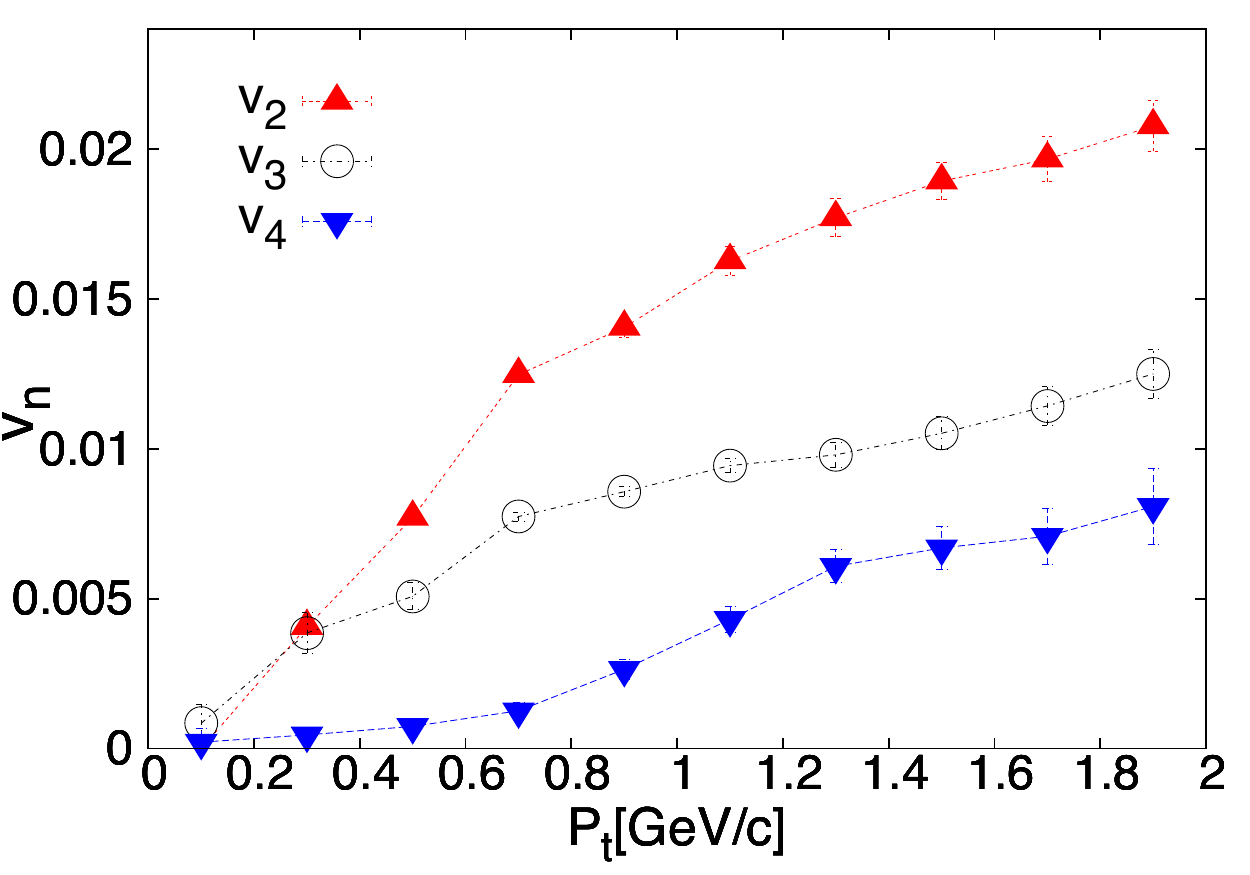}
\caption{\label{f:c2.5}%
Anisotropy parameters $v_n$ from centrality class 0--2.5\% for charged hadrons 
as functions of $p_t$. The energy  loss of hard 
partons is given by $dE/dx|_0 = 4$~GeV/fm.  Red $\triangle$: $v_2$,
black  $\circ$: $v_3$, blue $\nabla$: $v_4$.
}
\end{figure}

\begin{figure}[t]
\includegraphics[width=0.44\textwidth]{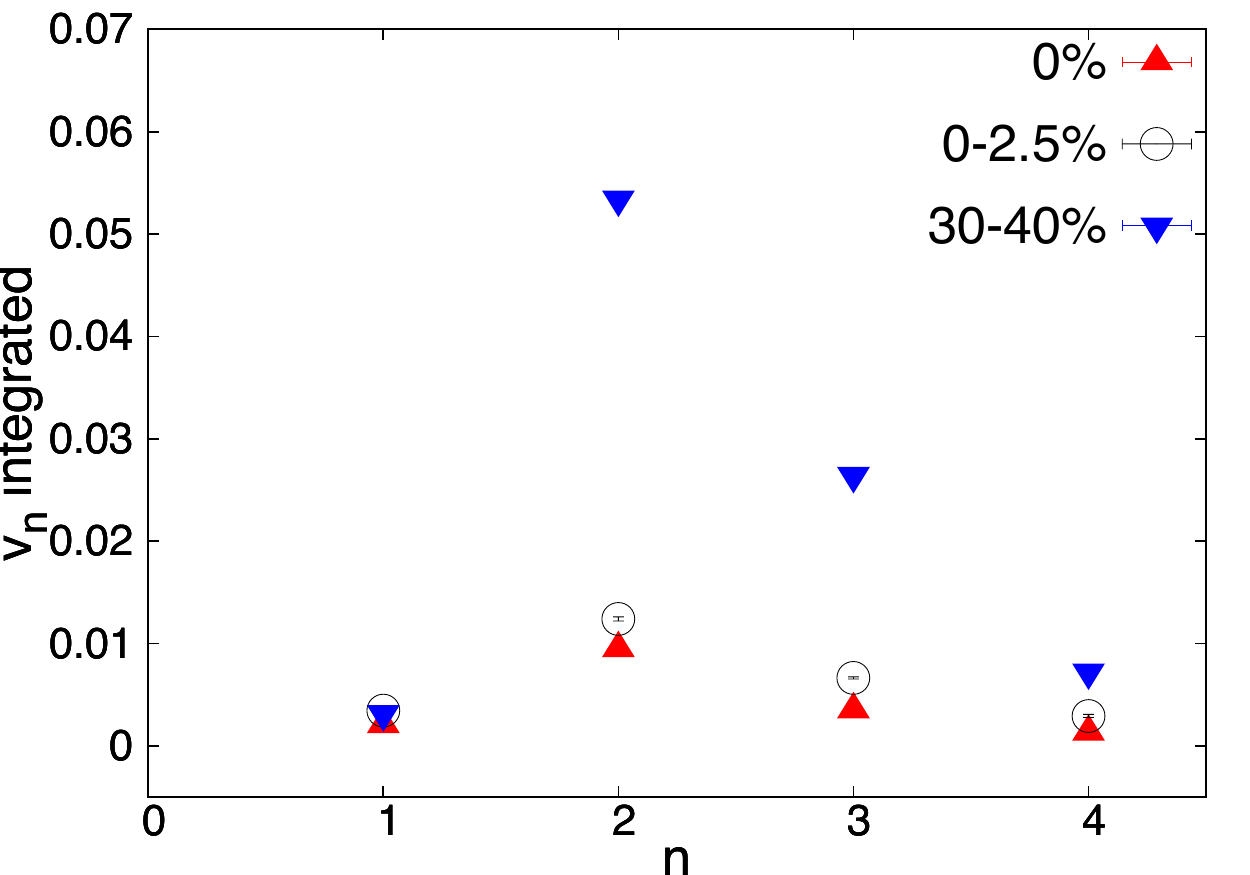}
\caption{\label{f:vnint}
Anisotropy parameters $v_n$ for charged hadrons integrated over $p_t$
for different centralities. The energy  loss of hard 
partons is given by $dE/dx|_0 = 4$~GeV/fm. 
}
\end{figure}

The CMS collaboration has found a strong dependence of $v_2$ and $v_3$ 
on centrality even for central collisions \cite{CMSvn}. Although here we only want to get an 
educated estimate on the size of the effect that our mechanism can generate, it is 
tempting now to look how our $v_n$'s would change if we go to centrality class 
0--2.5\%. The results are shown in Fig.~\ref{f:c2.5} for charged particles.
%%%%%%%%%%%%%%%%%%%%%%%%%%%%%%%%%%%%%%%%%%%%%%%%%%
%%%%%%%%%%%%%%%%%%%%%%%%%%%%%%%%%%%%%%%%%%%%%%%%%%%
In Fig.~\ref{f:vnint} we present the integarted $v_n$'s as functions of centrality. 
%%%%%%%%%%%%%%%%%%%%%%%%%%%%%%%%%%%%%%%%%%%%%%%%%%
%%%%%%%%%%%%%%%%%%%%%%%%%%%%%%%%%%%%%%%%%%%%%%%%%%%
We see that going from $b=0$~fm to 0--2.5\% centrality there is no dramatic 
increase in $v_n$'s. If such effect is present in data, it must be caused by a different 
mechanism.

\begin{figure}[t]
\includegraphics[width=0.44\textwidth]{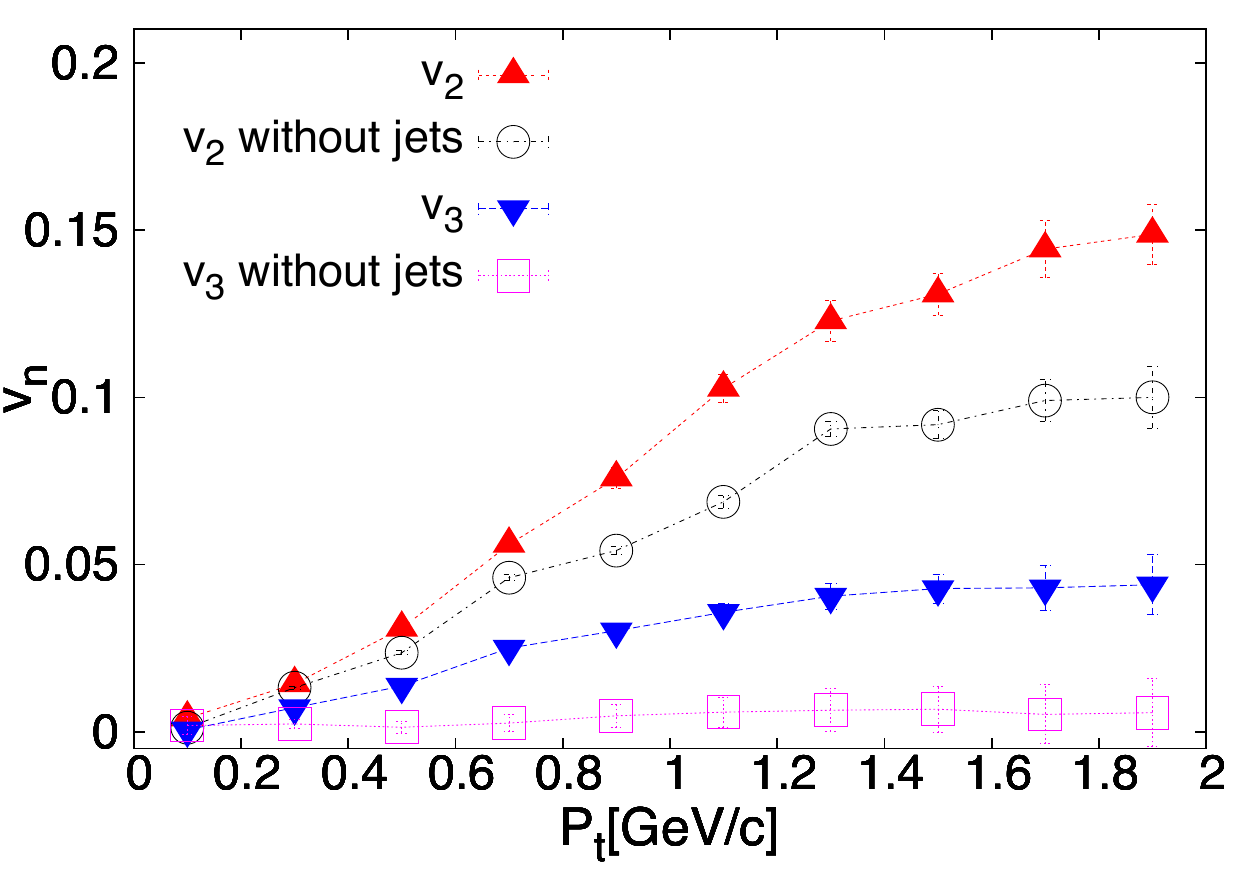}
\caption{\label{f:noncen}%
Anisotropy parameters $v_2$ and $v_3$ for charged hadrons 
as functions of $p_t$ from collisions within centrality class 30--40\%. 
The energy  loss of hard 
partons is given by $dE/dx|_0 = 4~$GeV/fm.
}
\end{figure}

In simulations of non-central events we clearly establish that the flow anisotropy generated 
by hard partons is correlated with the reaction plane. This is a consequence 
of the mechanism where two streams of the fluid in the wakes merge when they are 
close. Then they continue flowing in direction given by momenta of the two streams
\cite{Tomasik:2011xn,schulc}. The proof of validity of this mechanism is presented in Fig.~\ref{f:noncen}.
%%%%%%%%%%%%%%%%%%%%%%%%%%%%%%%%%%%%%%%%%%%%%%%%%%
%%%%%%%%%%%%%%%%%%%%%%%%%%%%%%%%%%%%%%%%%%%%%%%%%%%
We show $v_2$ and $v_3$
of charged hadrons as calculated from an ensemble of 500 events with 
hard partons depositing momentum.
They are  compared with $v_2$ and $v_3$ being only 
due to event-averaged almond shape of the initial hot matter. Obviously, $v_3$ 
must vanish then and it indeed does. If the contribution 
of hard partons had random direction, we would not expect an increase of $v_2$. 
However, $v_2$ increases by more than factor of 1.5. 

Note also the increase of other orders of the anisotropy presented in 
Fig.~\ref{f:vnint} for integrated $v_n$'s. 

Our results show that the interplay of many minijet-induced streams in a single 
nuclear collision at the LHC yields considerable contribution to azimuthal anisotropies 
of hadron distributions. The present simple non-viscous model with smooth 
initial conditions should merely be used for an educated estimate  
of the influence. It is certainly not capable of reproducing data, since this requires 
inclusion of many fine details. Among them the most prominent are shear and bulk 
viscosities and a tuned model of fluctuating initial conditions. It must be investigated, 
how to disentangle various mechanisms that generate all kinds of azimuthal 
anisotropies with the help of many features of data that are currently being measured. 

\begin{acknowledgments}
This work was supported in parts by APVV grant 0050-11, VEGA grant 
1/0457/12 (Slovakia) and 
M\v{S}MT grant  LG13031 (Czech Republic).
\end{acknowledgments} 

%%%%%%%%%%%%%%%%%%%%%%%%%%%%%%%%%%%%%%%%%%%%%%%%%%%%%

\end{document}